# Magneto-optical spaser

D. G. Baranov,[1,2] A. P. Vinogradov,[1,2] A. A. Lisyansky,[3,*] Yakov M. Strelniker,[4] and David J. Bergman[5]

[1]*Moscow Institute of Physics and Technology, 9 Institutskiy per., Dolgoprudniy 141701, Moscow Reg., Russia*
[2]*Institute for Theoretical and Applied Electromagnetics, 13 Izhorskaya, Moscow 125412, Russia*
[3]*Department of Physics, Queens College of the City University of New York, Queens, NY 11367, USA*
[4]*Department of Physics, Bar-Ilan University, IL-52900 Ramat-Gan, Israel*
[5]*Raymond and Beverly Sackler School of Physics and Astronomy, Faculty of Exact Sciences, Tel Aviv University, IL-69978 Tel Aviv, Israel*
*Corresponding author: lisyansky@qc.edu

We present an electrodynamical model of a novel quantum plasmonic device – magneto-optical (MO) spaser. It is shown that a spherical gain nanoparticle coated with a metallic MO shell can operate as a spaser amplifying circularly polarized surface plasmons. The MO spaser may be used in design of optical isolator in plasmonic transmission lines as well as in spaser spectrometry of chiral molecules.

Recently, nanosources of coherent light have attracted significant attention [1]. After Ref. [2], different plasmonic-based light sources have been studied theoretically [3–5] and realized experimentally [6–8]. In most cases, such a light source is a pumped optical emitter (a quantum dot or an active molecule) placed inside or near a plasmonic resonator. For the case of a metallic nanoparticle, such a coupled system represents a SPASER, whose properties have been studied in details during the last decade [2, 9–12]. However, to the best of our knowledge, radiation from all suggested sources is linearly polarized.

Nowadays, interaction of circularly polarized light with matter is becoming an attractive field of research [13–17]. The increasing interest in this topic stems from the fact that sources of circularly polarized light find a number of intriguing applications in molecular sensing [15, 16] and quantum information [17]. Sources most commonly used to generate circularly polarized light are either vertical cavity surface-emitting lasers or spin-polarized diodes [14]. The major feature of such sources is that they are diffraction limited.

In the present paper we suggest a MO spaser – a subdiffractive source of circularly polarized light operating in the near field. Employing the description of the gain medium in terms of negative losses, we study its spasing modes. We establish the spaser generation condition and show that for some values of pump intensity only one of two modes can realize.

A suggested MO spaser consists of an amplifying core, e.g. a quantum dot or active molecules, coated by a metallic layer exhibiting MO response at optical frequencies (see a schematic drawing in Fig. 1a). The whole core-shell nanoparticle is of subwavelength size. The dielectric permittivity of gain medium is approximated by the Lorentzian profile with "negative" losses [18, 19]. The simplest expression for the permittivity suitable for our purpose may be deduced from the Maxwell-Bloch equations [20]. In the framework of this approach, the evolution of electric field **E** is related to the macroscopic polarization **P** of a gain subsystem via classical Maxwell's equation, while the dynamics of the polarization and the population inversion of active atoms $n$ is governed by the equations following from the density matrix formalism (see [18] for detailed derivation). The gain atoms embedded into the host medium are modeled as two-level systems with transition dipole moment **μ** spread in the host matrix. Assuming harmonic time dependence of the electric field, $e^{-i\omega t}$, we obtain the relation between the polarization $P$ and the electric field $E$ inside the medium, resulting in the following expression for nonlinear permittivity of a gain medium with the anti-Lorentzian profile:

$$\varepsilon_{gain}(\omega) = \varepsilon_0 + D_0 \frac{\omega_0}{\omega} \frac{-i + \frac{\omega^2 - \omega_0^2}{2\omega\Gamma}}{1 + \beta|E(\omega)|^2 + \left(\frac{\omega^2 - \omega_0^2}{2\omega\Gamma}\right)^2}, \qquad (1)$$

where $D_0 = 4\pi\mu^2\tau_p n_0 / \hbar$ describes the population inversion between excited and ground states of the active atoms, $\beta = \mu^2 \tau_n \tau_p / \hbar^2$ and $\Gamma = 1/\tau_p$. Here $\tau_p$ and $\tau_n$ are relaxation times for polarization and inversion, respectively, and $n_0$ stands for pumping of active atoms.

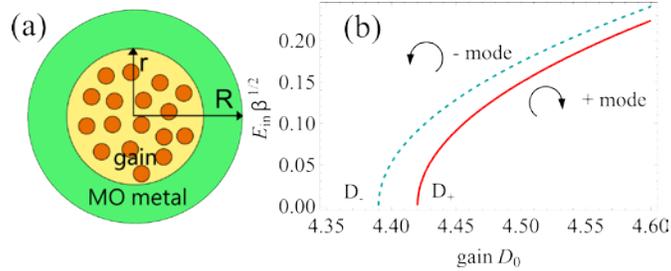

Fig. 1. (a) A schematic drawing of the suggested composite core-shell MO spaser. (b) Amplitudes of spasing modes of the MO spaser versus gain near the bifurcation point.

Dielectric properties of MO metal, which is assumed to be cobalt, are described with the permittivity tensor having off-diagonal elements accounting for the MO effect:

$$\hat{\varepsilon}_{MO} = \begin{pmatrix} \varepsilon & ig & 0 \\ -ig & \varepsilon & 0 \\ 0 & 0 & \varepsilon \end{pmatrix}. \tag{2}$$

For values of the diagonal and off-diagonal elements of the permittivity tensor, we use experimental data obtained in Ref. 21.

We start our consideration with eigenmode analysis of a MO spaser within the quasistatic approximation by finding steady-state solutions with no applied electric field. The key assumption we make here is that the electric field inside the nonlinear amplifying core is homogenous: $\mathbf{E}_{core}(\mathbf{r}) \equiv \mathbf{E}$. Though this may be not the *only* possibility, the rigorous consideration requires finding a solution of a nonlinear Laplace problem, which may lead to cumbersome calculations but would not change results qualitatively. Fortunatelly, this special geometry allows us to solve the nonlinear problem analytically and to obtain an exact solution. We write the field inside the metallic shell in the form $\mathbf{E}_{shell} = \hat{B}\mathbf{E} - \hat{C}\mathbf{E}/\rho^3 + 3(\hat{C}\mathbf{E},\mathbf{n})\mathbf{n}/\rho^3$ and the field outside nanoparticle as $\mathbf{E}_{ext} = -\hat{A}\mathbf{E}/\rho^3 + 3(\hat{A}\mathbf{E},\mathbf{n})\mathbf{n}/\rho^3$, where $\hat{A}, \hat{B}$ and $\hat{C}$ are some unknown tensors and $\rho$ is the distance from the center of the gain core. Boundary conditions lead us to the following system of equations:

$$\begin{cases} \varepsilon_{gain}(|E|)\mathbf{E} = \varepsilon_{shell}(\hat{B} + 2\hat{C}/r^3)\mathbf{E} + \hat{G}(\hat{B} - \hat{C}/r^3)\mathbf{E}, \\ \hat{1} = \hat{B} - \hat{C}/r^3, \\ \varepsilon_{shell}(\hat{B} + 2\hat{C}/R^3) + \hat{G}(\hat{B} - \hat{C}/R^3) = 2\hat{A}/R^3, \\ \hat{B} - \hat{C}/R^3 = -\hat{A}/R^3, \end{cases} \tag{3}$$

where $r$ and $R$ are radii of the gain core and the metallic shell, respectively (see Fig. 1a), and $\hat{G}$ stands for the non-diagonal part of the permittivity tensor of the MO shell (see Eq. 2). Obtaining explicit expressions for $\hat{A}$, $\hat{B}$ and $\hat{C}$ from the last three equations of system (3), we arrive at the eigenvalue problem for a combination of these tensors:

$$\varepsilon_{gain}(E,\omega)\mathbf{E} = \hat{M}\mathbf{E}, \tag{4}$$

where we defined $\hat{M} = \varepsilon_{shell}(\hat{B} + 2\hat{C}/r^3) + \hat{G}(\hat{B} - \hat{C}/r^3)$. This equation determines the possibility of existence of non-zero internal field inside the nanoparticle at zero incident field and amplitude of this field. Therefore, it can be treated as the condition for spasing. We only consider solutions with $E_z = 0$, for which the polarization is perpendicular to the magnetization vector of the MO shell. From the symmetry of the permittivity tensor it follows that two non-diagonal elements of $\hat{M}$ have opposite sings:

$$\hat{M} = \begin{pmatrix} M_{1,1} & M_{1,2} \\ -M_{1,2} & M_{2,2} \end{pmatrix} \qquad (5)$$

and therefore, in the subspace $E_z = 0$, its two eigenvalues are given by $\lambda_\pm = M_{1,1} \pm iM_{1,2}$ with two eigenvectors $\mathbf{E}_\pm = (1, \pm i, 0)$ corresponding to the right and left circular polarizations of the electric field inside the amplifying core. All the tensors involved ($\hat{A}, \hat{B}$ and $\hat{C}$) have the same anti-Hermitian structure in the $E_z = 0$ subspace as the dyadic $\hat{M}$. Thus, the condition for spasing can be recasted as

$$\varepsilon_{gain}(\omega, E_{in}, D_0) = \lambda_\pm, \qquad (6)$$

$$\lambda_\pm = \frac{(-2 + 2\varepsilon \mp g)(\varepsilon \pm g)r^3 - (2 + \varepsilon \pm g)(2\varepsilon \mp g)R^3}{(-2 + 2\varepsilon \mp g)r^3 + (2 + \varepsilon \pm g)R^3}. \qquad (7)$$

Equation (6) defines two spasing modes with circular polarizations. As it was shown in our previous work, their dependence on gain exhibits the Hopf bifurcation [22]. Indeed, the right hand side of Eq. (6) does not depend on $E_{in}$ and $D_0$ and hence, has a nonvanishing imaginary part for all frequencies in case of lossy metal. At the same time, $\varepsilon_{gain}$ has an arbitrary small imaginary part for vanishing $D_0$. Threshold gain for spasing $D_{th}$, which is the smallest possible gain $D_0$ for which Eq. (6) is satisfied, is achieved for $E_{in} = 0$.

To model the gain medium, we use values of parameters which are typical for organic dyes [23]: $\omega_0 = 2\,\mathrm{eV}$, $\varepsilon_0 = 2$, and $\gamma = 0.05\,\mathrm{eV}$. Electric field is measured in the units of $\beta^{-1/2}$, so its value is of no importance. The radii of gain core and MO shell are $r = 20\,\mathrm{nm}$ and $R = 1.1\,r$.

Our calculations show that the two spasing modes have different spasing frequencies $\omega_\pm$ and different spasing thresholds $D_\pm$. In Fig. 1b we depict the mode amplitudes (the field inside the amplifying core $E_{in}$) versus gain $D_0$ for a MO spaser. One can see that in the interval of gain values $4 < D_0 < 5$ the two spasing modes arise with the square-root-like dependence of amplitudes on gain. The calculated spasing frequencies are $\omega_- = 1.976\,\mathrm{eV}$ and $\omega_+ = 1.977\,\mathrm{eV}$.

Fig. 1b shows that there exists a range of gain $D_0$ for which only left or right circularly polarized spasing mode is emitted. Which particular mode is realized depends on the direction of the magnetization vector. The difference between thresholds of the modes is simply the consequence of the circular dichroism of the magneto-optical shell. Indeed, in the circular polarization basis the permittivity tensor of MO metal is $\hat{\varepsilon}_{MO} = \mathrm{diag}[\varepsilon + g, \varepsilon - g, \varepsilon]$, so that $\varepsilon + g$ eigenvalue corresponds to $\mathbf{E}_- = (1, -i, 0)$ polarization and $\varepsilon - g$ to $\mathbf{E}_+ = (1, i, 0)$, respectively. This means that effectively permittivity of MO metal differs for right and left polarized modes leading to different $Q$-factors of these modes. As a result, the threshold for one of the two modes is smaller.

For the core-shell spaser considered above, the developed approach predicts the value of threshold gain as $D_{th} \simeq 4.5$ leading to the bulk gain coefficient $k'' \simeq D_{th}\omega/(2c) \simeq 10^5$ cm$^{-1}$. This value is unachievable for typical organic dyes [23] even for full population inversion. Fortunately, the described effect can be observed in systems using plasmonic metals and dielectric low-loss MO materials, such as Yittrium Iron Garnet (YIG).

To overcome large ohmic damping and unrealistic gain required for operation of the Co-based spaser, we suggest two modifications of the original design: a spherical gain core coated first with a silver shell and on top of it with another shell of magneto-optical cobalt or YIG (Fig. 2a). Due to the presence of silver shell, such a structure forms a high-$Q$ plasmonic resonator exhibiting MO properties. The amplitudes of the two circularly-polarized modes are governed by the same Eq. (6) in which modified $\lambda_\pm$ are now defined by boundary conditions at three interfaces $r_1 < r_2 < R$.

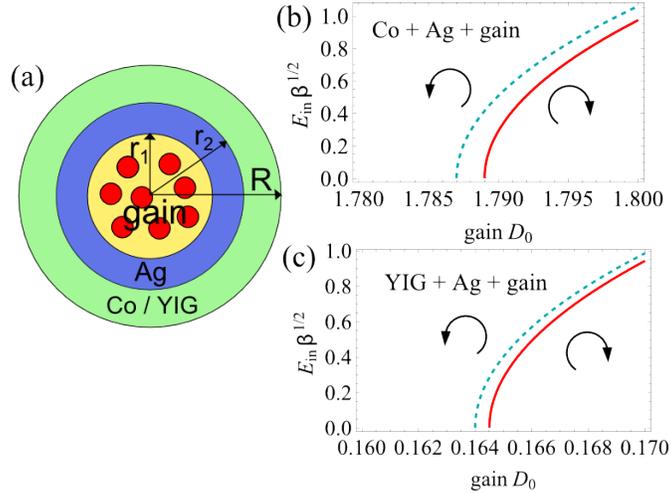

Fig. 2. (a) Alternative design of the MO spaser allowing for low-threshold spasing. (b) and (c) amplitudes of spasing modes versus gain for Co + Ag and YIG + Ag spaser respectively.

The threshold gain of these structures depends on the radii of the gain core and silver and MO shells and background permittivity of gain medium $\varepsilon_0$. Gain required for spasing can be minimized by choosing optimal parameters for both structures. In addition, one should keep in mind that the plasmon resonance of core/shell nanoparticles has to be tuned to the frequency range of the MO resonance of MO material. The calculations show that threshold for Co/Ag spaser can be as low as $D_{th} \approx 1.8$ for $r_1 = 0.95R$, $r_2 = 0.6R$ and $\varepsilon_0 = 5$. When the thickness of the cobalt layer is larger than $0.005R$, spasing threshold quickly increases almost to its original unrealistic value $D_{th} \approx 3 \div 4$. For the case of YIG/Ag spaser, the conditions for spasing are even more favorable: by choosing $r_1 = 0.9R$, $r_2 = 0.7R$ and $\varepsilon_0 = 2$ we attain decrease of the threshold up to $D_{th} = 0.17$ leading to the bulk gain coefficient of approximately $k'' \simeq 3 \times 10^3 \text{ cm}^{-1}$, which is achievable for organic amplifying media.

In conclusion, we have suggested a MO spaser representing a novel ultra-small source of coherent circularly polarized light. Unlike a non-magnetic spaser, this light source has two spasing modes with left and right cirlcular poalarizations of the dipole moments. Each of the modes has different pumping thresholds so that for a certain range of values of pumping only a mode with a specific circular polarization realizes. This is especially important for prospective applications for quantum information, where manipulations of qubits require illumination with a circularly polarized light.